\documentstyle[11pt]{article}
\begin{document}
\title{Gauge Fixing in the Chain by Chain Method}
\author{A. Shirzad\thanks{shirzad@cc.iut.ac.ir} \hspace{1cm} F. Loran\thanks{loran@cc.iut.ac.ir}\\ \\
  {\it Department of  Physics, Isfahan University of Technology (IUT)}\\
{\it Isfahan,  Iran,} \\
  {\it Institute for Studies in Theoretical Physics and Mathematics (IPM)}\\
{\it P. O. Box: 19395-5531, Tehran, Iran.}}
\date{}
\maketitle

\begin{abstract}In a recent work we showed that for a Hamiltonian
system with constraints, the set of constraints can be
investigated in first and second class constraint chains. We show
here that using this "chain by chain" method for an arbitrary
system one can fix the gauges in the most economical and
consistent way. We show that it is enough to assume some gauge
fixing conditions conjugate to last elements of first class
chains. The remaining necessary conditions would emerge from
consistency conditions.
\end{abstract}

\section{Introduction} It is well known that gauge theories correspond to
Hamiltonian constraint systems with first class constraints. Dirac
has conjectured that first class constraints (primary or
secondary) are generators of gauge transformations \cite{Dirac}.
Despite some counterexamples \cite{Cow} one can assume the
validity of Dirac conjecture under suitable regularity conditions
\cite{Henbook}. The presence of first class constraints and the
associated gauge freedoms indicates that corresponding to any
given physical state there exist some orbit in phase space, i. e.
gauge orbit. Gauge transformations translate the system along
gauge orbits. One can impose further restrictions on the canonical
variables, gauge fixing conditions, to make a one to one
correspondence between them and physical states. In this way the
initial phase space reduces to a smaller one on which both
constraints and gauge fixing conditions (GFC) do vanish. This
subspace is called the reduced phase space. There are three
properties that a satisfactory set of constraints and GFC's should
satisfy:
\par
\ \ $i)$ The set of constraints should be regular and irreducible
\cite{Henbook}.
\par
\ $ii)$ The GFC's should be accessible. They should intersect the
gauge orbits at least once. In addition they should completely fix
the gauges.
\par
$iii)$ The GFC's should remain valid during the time i.e. their
time derivatives should vanish.
\par
The property $(ii)$ is well known. The first and third properties
though considered practically\footnote{See for example
ref.\cite{Gov}.}, but are not emphasized through the literature.
In most cases, people work with well-behaved models possessing
regular and irreducible constraints and propose suitable GFC's
considering the second property mentioned above.\par Since first
class constraints are generators of gauge transformation the
process of gauge fixing strongly depends on the method of
producing the constraints. For example, there are some methods
which lead to a set of reducible constraints
\cite{ShirShab,CaboChai}. In these cases one needs primarily a
consistent method to distinguish the independent gauge degrees of
freedom.

In ref. \cite{Chain} we proposed a new method, the chain by chain
method, for constructing an irreducible set of constraints. In
this method constraints are classified in a number of second class
and a number of first class constraint chains. This article is
devoted to gauge fixing in the chain by chain method. We show that
one only needs to find GFC's that fix the gauge freedoms
associated to the last element of first class chains. Consistency
conditions generate the remaining needed GFC's. In this way the
properties $(i)-(iii)$ are satisfied consistently. Moreover, the
number of necessary GFC's to be found is just equal to the number
of first class chains that in general is less than the number of
first class constraints. We do not consider difficulties due to
Gribov ambiguities \cite{Gribov} and the problem of covariance of
the formalism in this work.
\par In the following section we review basic concepts of
constraint systems and gauge transformations in the extended and
total Hamiltonian formalism. The chain by chain method is also
reviewed briefly in that section. Our method for gauge fixing in
the framework of chain by chain method is proposed in section 3.
In section 4 we examine our method in Electrodynamics and
Yang-Mills theories. Some concluding remarks are given in section
5.
\section{Constraints and Gauges}
Consider a dynamical system given by a canonical Hamiltonian
$H_c(q,p)$ and a set of primary constraints $\phi_1^a(q,p)$,
$a=1,\ldots,n$. The Hamilton-Dirac equations of motion for an
arbitrary function $g(q,p)$ read \cite{Dirac}
\begin{equation}\label{a1}
\dot{g}(q,p)=\left\{g,H_T\right\},
\end{equation}
where
\begin{equation}\label{a2}
H_T=H_c+\sum_a v_a\phi^a_1,
\end{equation}
in which, $v_a$ are Lagrange multipliers. Equation (\ref{a1})
together with constraint relations $\phi_1^a(q,p)=0$ can be
derived by varying the total action
\begin{equation}
S_T=\int dt \left(\dot{q}_ip_i-H_T\right),
\end{equation}
with respect to canonical variables $(q_i,p_i)$ and Lagrange
multipliers $v_a$. Gauge transformations are defined as
transformation on phase space trajectories $(q_i(t),p_i(t))$ and
Lagrange multipliers that include arbitrary functions of time and
leave the total action $S_T$ invariant. In models satisfying Dirac
conjecture one can show that gauge transformations transform
different classes of solutions, belonging to different choices of
arbitrary functions of time, to each other \cite{Henbook}.
\par As is well known consistency conditions for primary
constraints, $\dot{\phi}^a_1=0$, may lead to determination of some
Lagrange multipliers or appearing secondary constraints. In the
traditional method of producing the secondary constraints, i.e.
the level by level method
\cite{Henbook,BatGom,HenTeiZan,GoHenPo,GraciaPon}, constraints
appear in a sequence of levels of irreducible constraints. The
primary constraints form the first level. One obtains the
constraints of the $n$-th level,say, by considering the
consistency of constraints of the $(n-1)$-th level. By
construction no new constraint emerges from the consistency
conditions of the last level.

 In the chain by chain method, conversely,
\cite{Chain} one investigates the consistency of primary
constraints one by one. For primary constraint $\phi^a_1$, say,
the corresponding chain is knitted via the recursion relation
\begin{equation}\label{a3}
\phi_n^a=\left\{\phi_{n-1}^a,H_c\right\}.
\end{equation}
Some chains terminate when a Lagrange multiplier is determined.
These are second class chains that contain only second class
constraints. The remaining chains, first class chains, which
contain only first class constraints, end up when consistency of
the last element is achieved identically. The whole algorithm is
given in \cite{Chain}. Following this algorithm one can separate
first class and second class constraints from each other and
arrange them in the associated chains. In addition constraints in
different chains commute with each other, i.e. the Poisson bracket
of any element of one chain with any element of other chains
vanishes on the surface of the constraints. Therefore the
structure of first class chains do not change if one replaces the
Poisson brackets with Dirac brackets and eliminates the second
class constraints. Consequently one can consider every constraint
system as a purely first class system when the question of gauge
fixing arises. In the following we study gauge fixing in first
class systems. The above observations guarantee the validity of
our results in general cases.
\section{Gauge Fixing}
Consider a system with $N$ first class constraints arranged in $m$
first class chains:
\begin{equation}
\begin{array}{cccccc}
\phi^1_1 & \phi^2_1 & \ldots & \phi^a_1 & \ldots & \phi^m_1\\
\vdots   & \vdots   &        &  \vdots  &        &\vdots   \\
         &          &        &          &        & \phi^m_{N_m}\\
\vdots &\phi^2_{N_2}&        & \vdots     \\
\phi^1_{N_1}\\
\        &          &         & \phi^a_{N_a}
\end{array}\label{a48}\end{equation}
The evolution of gauge invariant quantities may also be determined
by the extended Hamiltonian
\begin{equation}
H_E=H_c+\sum_{a,i}\lambda^a_i\phi^a_i\label{d1},
\end{equation}
where $\lambda^a_i$ are undetermined Lagrange multipliers, which
here can be considered as independent gauge parameters. In the
extended formalism, corresponding to each first class constraint
there exist one Lagrange multiplier to be determined by gauge
fixing. Therefore one should impose an equal number of independent
gauge fixing conditions as there are first class constraints. The
consistency of gauge fixing conditions determines the Lagrange
multipliers. The true dynamics of a constrained system, however,
is given by the total Hamiltonian defined in Eq.(\ref{a2}). The
extended Hamiltonian can be used instead of the total Hamiltonian,
provided that one demand after all that Lagrange multipliers
corresponding to secondary constraints (and their variations)
vanish \cite{Henbook}.

For several reasons gauge fixing in the total Hamiltonian
formalism requires some care. First, the number of gauges to be
fixed is $N=\sum_{a=1}^m N_a$, the total number of first class
constraints; while the number of Lagrange multipliers to be
determined is $m$, which is usually less than $N$. Second, the
consistency of GFC's may lead to additional constraints that
over-determine the system. Third, the (first class) constraints in
the total Hamiltonian formalism do not generate independent gauge
transformations. It can be shown \cite{ShirShab,CaboChai} that
there exist $(N-m)$ differential equations among the gauge
parameters corresponding to first class constraints. The question
arises that "how can one fix the independent gauges in a
consistent way?". This can be answered within the framework of the
chain by chain method in a simple way as follows.
\par Considering the set of first class constraints given in
Eq.(\ref{a48}), one may find $m$ gauge fixing conditions
$\Omega^{a}_{N_a}$'s with the following property:
\begin{equation}
\{\Omega^{a}_{N_a},\phi^b_n\} \approx \eta^a(q,p) \delta^{ab}
\delta_{n,{N_a}} \label{a49}\end{equation} where $\eta^a(q,p)$ are
some arbitrary functions which should not vanish on the surface of
the constraints. In principle the set of first class constraints
$\phi^a_{N_a}$'s can be considered as a set of momenta. In such an
idealized system the gauge fixing conditions $\Omega^a_{N_a}$'s
are the corresponding conjugate coordinates and consequently
$\eta^a$'s become proportional to the unity. Therefore, the
existence of $\eta^a$'s can always be assumed. \par We show that
the remaining GFC's needed to fix the gauge completely can be
obtained by using the consistency of $\Omega^a_{N_a}$'s. Since
$\{\Omega^a_{N_a},\phi^b_1\} \approx 0$, the consistency of
$\Omega_{N_a}$'s i.e. $\dot{\Omega}_{N_a}=0$, gives  a new set of
GFC's as:
\begin{equation}
\Omega^a_{N_a-1} \equiv \{\Omega^a_{N_a},H_c\}.
\label{a50}\end{equation} Let us consider the Poisson bracket of
$\Omega^a_{N_a-1}$ with the constraints:
\begin{eqnarray}
\{\Omega^a_{N_a-1},\phi^b_n\}&=&\{\{\Omega^a_{N_a},H_c\},\phi^b_c\}\nonumber \\
 &=& \{H_c,\{\phi^b_n,\Omega^a_{N_a}\}\}-\{\Omega^a_{N_a},\phi^b_{n+1}\}
\label{a51}\end{eqnarray} where we have used Eq.(\ref{a3}) in the
last line. Using Eq.(\ref{a49}) the above expression vanishes for
$a\ne b$, as well as for $a=b$ and $n<N_a-1$. Note specially that
the Poisson brackets of $\Omega^a_{N_a-1}$ with the primary
constraints vanishes. For $a=b$ and $n=N_a-1$ Eq.(\ref{a51})
gives:
\begin{equation}
\{\Omega^a_{N_a-1},\phi^a_{N_a-1}\}\approx -\eta^a(q,p).
\label{a52}\end{equation} Consistency of $\Omega^a_{N_a-1}$ leads
to $\Omega^a_{N_a-2}\equiv \{\Omega^a_{N_a-1},H_c\}$ and so on.
The generic terms for the GFC's are related to each other as
follows:
\begin{equation}
\Omega^a_n=\{\Omega^a_{n+1},H_c\},\ \ \ \ \ \ n=1,\ldots,N_a-1
\label{a53}\end{equation} Comparing Eq.(\ref{a53}) with
Eq.(\ref{a3}) one realizes that the chains of GFC's are exactly
the "mirror images" of the constraint chains, i.e. they are
knitted in the opposite direction. The whole story goes on as
follows: one begins with $\phi^a_1$, goes through consistency
conditions until reaches $\phi^a_{N_a}$, then fixes the gauge by
finding $\Omega^a_{N_a}$ conjugate to $\phi^a_{N_a}$, turns all
the way round through consistency conditions to reach $\Omega^a_1$
at the end point. The story sounds more interesting by repeating
the calculations given in Eq.(\ref{a51}) to get:
\begin{equation}
\begin{array}{l}
\{\Omega^a_n,\phi^b_{n'}\}\approx 0\ \ \ \ \ a \ne b \\
\{\Omega^a_n,\phi^a_{n'}\}\approx 0\ \ \ \ \ n'<n \\
\{\Omega^a_n,\phi^a_n\}\approx(-1)^{N_a-n}\eta^a(q,p)
\end{array}\label{a54}\end{equation}
As is observed each $\Omega^a_n$ is really conjugate to its
partner $\phi^a_n$. The story ends when one investigates the
consistency of $\Omega^a_1$'s where the Lagrange multipliers are
determined due to non-vanishing Poisson brackets
$$\{\Omega^a_1,\phi^a_1\}=(-1)^{N_a-1}\eta^a(q,p).$$
Using Eqs.(\ref{a54}) the matrix of Poisson brackets of
constraints with GFC's can be obtained as follows:
\begin{equation}
\left(\begin{array}{cccc}
 {\left(\begin{array}{ccc}
  \eta^1&         & 0   \\
  \ddots &\ddots\\
         \ddots &\ddots &e_1\eta^1
 \end{array}\right)}&&&\mbox{\Huge 0}\\
 &{\left(\begin{array}{ccc}
  \eta^2 &       & 0   \\
  \ddots & \ddots \\
         \ddots & \ddots &e_2\eta^2
 \end{array}\right)}\\
 &&\ddots \\
 \mbox{\Huge 0}&&&{\left(\begin{array}{cccc}
  \eta^m &        & 0 \\
  \ddots & \ddots \\
         \ddots & \ddots &e_m\eta^m
 \end{array}\right)}
\end{array}\right)\label{a55}\end{equation}
where $e_a=(-1)^{N_a-1}$. As is obvious, the determinant of the
matrix given in (\ref{a55}) is proportional to $\prod_a
[\eta^a(q,p)]^{N_a} \neq 0$. Since chain by chain method
guarantees that the set of first class constraints $\phi^a_n$'s is
irreducible this result ensures that the above $\Omega^a_n$,
completely fix the gauges \cite{Henbook}. Each non-vanishing block
in the matrix (\ref{a55}) corresponds to a definite constraint
chain. There emerge indeed some non-vanishing elements below the
diameter coming from $\{\Omega^a_n,\phi^a_{n'}\}$ with ${n'}>n$.
One can redefine constraints and GFCs properly to make these off
diagonal elements vanish (see \cite{Chain})

\section{Electrodynamics with source and Yang-Mills}
As a first example of applying the method let us consider
electrodynamics with bosonic source given by the Lagrangian:
\begin{equation}
L=\int d^3x \{-\frac {1}{4} F_{\mu \nu}F^{\mu \nu}-\frac
{1}{2}\bracevert(\partial _{\mu} +igA_\mu )\Phi \bracevert^2-
v(\Phi\Phi^*)\} \label{a56}\end{equation} where $V(\Phi\Phi^*)$ is
a potential and
\begin{equation}
F_{\mu \nu}=\partial_\mu A_\nu -\partial_\nu A_\mu.
\label{a57}\end{equation} Rewriting $L$ in terms of the dynamical
fields $A^\mu (x,t)$, $\eta (x,t)$ and $\psi (x,t)$ where
\begin{equation}
\Phi (x,t)=\eta (x,t)e^{i\psi (x,t)}, \label{a58}\end{equation}
the canonical momenta are
\begin{equation}
\Pi ^\mu =-F^{0\mu},\hspace{1cm}\pi _\eta
=\dot{\eta},\hspace{1cm}\pi _\psi =\eta^2(\dot{\psi}+gA_0).
\label{a59}\end{equation} It is obvious from Eq.(\ref{a57}) that
$\phi _1=\Pi _0$ is our primary constraint. Then the total
Hamiltonian can be written as
\begin{equation}
\begin{array}{l}
H_T=\int d^3x\{{\cal H}^{ED}+\frac {1}{2}\pi ^2_\eta +\frac
{1}{2\eta ^2}
\pi^2_\psi -gA_0\pi_\psi\vspace{2mm}\\
\hspace{1cm}+\frac {1}{2} \eta ^2(\partial _k\psi )(\partial_k
\psi ) +\frac {1}{2} (\partial _k\eta)(\partial _k\eta)+g\eta
^2A_k(\partial _k\psi+\frac {1}{2}gA_k)
\vspace{2mm}\\
\hspace{1cm}+V(\eta )+v(x,t)\Pi ^0(x,t)\}
\end{array}\label{a60}\end{equation}
where $v(x,t)$ is the Lagrange multiplier (field) and
\begin{equation}
{\cal H}^{ED}=\frac{1}{2} \Pi _i\Pi _i+\frac{1}{4}
F_{ij}F_{ij}-A_0\partial _i\Pi _i. \label{a61}\end{equation} We
have ignored a surface term in Eq.(\ref{a61}) due to boundary
conditions. The secondary constraint serves as
\begin{equation}
\phi_2=\{\Pi^0,H_T\}=\partial _i\Pi_i+g\pi_\psi.
\label{a62}\end{equation} No further constraints emerges since
$\{\phi_2,H_T\}=0$. There is just one constraint chain with two
elements.
\par
To fix the gauge one should begin with $\Omega _2$ conjugate to
$\phi_2$. A simple choice is the Coulomb gauge $\Omega _2=\partial
_iA_i$. Consistency condition of $\Omega _2$ then gives another
GFC as
\begin{equation}
\Omega _1=\{\Omega _2,H_c\}=\partial _i\Pi_i +\partial _i\partial
_i A_0. \label{a63}\end{equation} Using Eq.(\ref{a62}) one has
$\partial _i\Pi_i\approx g\pi_\psi$, hence from Eq.(\ref{a63}) the
scalar potential $A^0$ is determined in this gauge to be
\begin{equation}
A^0(x,t)=\int d^3y\frac{g\pi_\psi (y,t)}{\bracevert
x-y\bracevert}. \label{a64}\end{equation} One important point to
be noted is that if one has imposed the famous gauges $\Omega
_2=\partial _iA_i$ and $\Omega _1=A_0$ then the consistency
condition $\dot{\Omega}_2=0$ would over-determine the system by
imposing $\pi _\psi =0$.
\par As a second example consider pure Yang-Mills
theory given by:
\begin{equation}
L=-\frac{1}{4}\int d^3x \mbox{Tr} (F_{\mu \nu}F^{\mu \nu})
\label{a67}\end{equation} where
\begin{equation}
F^{\mu \nu}=\partial ^\mu A^\nu -\partial ^\nu A^\mu
+ig[\Lambda^\mu ,\Lambda^\nu] \label{a68}\end{equation} The
dynamical fields $A^a_\mu(x,t)$ are implemented as
\begin{equation}
A^\mu =\Lambda ^\mu _a \Lambda^a \label{a69}\end{equation} where
$\Lambda _a$'s are generators of a Lie algebra with structure
constants $C_{ab}^c$:
\begin{equation}
[\Lambda ^a,\Lambda^b]=iC^{ab}_c\Lambda ^c.
\label{a70}\end{equation} The canonical momenta are $\Pi^a_\mu
=-F^a_{0\mu}$, where $\phi ^a_1=-\Pi^a_0$ serves as the set of
primary constraints. The canonical Hamiltonian is
\begin{equation}
H_c=\int d^3x\{\frac{1}{2}\Pi ^a_i \Pi ^a_i-A^a_0\partial
_i\Pi^a_i+
gA^a_0A^b_0C^{ab}_c\Pi^c_i+\frac{1}{4}F^a_{ij}F^a_{ij}\}
\label{a71}\end{equation} where a surface term is ignored. The
total Hamiltonian is
\begin{equation}
H_T=H_c+\int d^3x v^a(x,t)\Pi ^a_0(x,t). \label{a72}\end{equation}
The secondary constraints follow from the consistency of primary
constraints as:
\begin{equation}
\phi^a_2(x,t)=\{\Pi ^a_0,H_T\}\approx \partial
_i\Pi^a_i-gA^b_iC^a_{bc}\Pi^c_i. \label{a73}\end{equation} As in
electrodynamics, one may choose the first set of GFC's as
\begin{equation}
\Omega ^a_2=\partial _iA^a_i\approx 0. \label{a74}\end{equation}
Consistency of this gauge leads to
\begin{equation}
\Omega ^a_1\equiv \{\Omega ^a_2,H_c\}\approx \partial _i\Pi
^a_i+M^a_b A^b_0\approx 0 \label{a75}\end{equation} where
\begin{equation}
M^a_b=\delta ^a_b \partial _i\partial _i+gC^a_{bc}A^c_i\partial
_i. \label{a76}\end{equation} To see what is the consequence of
imposing the GFC's $\Omega^a_1\approx 0$ on $A^a_0$'s, let define
the Green function $G^b_c$  due to operator $M^a_b$:
\begin{equation}
M^a_b(x)G^b_c(x,y)=\delta ^a_c \delta(x-y).
\label{a77}\end{equation}  Eq.(\ref{a75}) can be solved:
\begin{equation}
A^a_0(x,t)=-\int d^3y \partial _i\Pi ^b_i(y,t)G^a_b(x,y)=H^a(x,t).
\label{a78}\end{equation} We observe again that the famous gauges
$A^a_0\approx 0$ and $\partial _iA^a_i\approx 0$, over-determine
the system by imposing an additional condition $H^a(x,t)\approx
0$.
\section{Conclusion}
Chain by chain method provides a simple constraint structure. In
this method the constraints are irreducible. Each Constraint
belongs to a chain that is identified by one of the primary
constraints. Some chains possess only second class and others
possess only first class constraints. Constraints in different
chains have vanishing Poisson brackets and constraints belonging
to each chain satisfy the recursion relation given in
Eq.(\ref{a3}). This structure provides a simple and consistent
method for gauge fixing. One searches for a set of constraints
that eliminate the gauge freedom associated to the last elements
of first class chains. One obtains the remaining necessary gauge
fixing conditions by imposing consistency conditions. In this
method gauge freedom associated to first class constraints
belonging to each first class chain is fixed indeed by only one
gauge fixing condition. This is exactly the case in the Lagrangian
formalism. Given a Lagrangian, one may fix the gauge partly by
adding some proper terms to the Lagrangian. Switching to the
Hamiltonian formalism the corresponding primary first class
constraints disappear and consequently the related first class
chains would not emerge. In other words every gauge fixing term
that is added to the Lagrangian fixes the gauge freedom associated
to one first class chain. This confirms our method for gauge
fixing.

\end{document}